# Germanene: a novel two-dimensional Germanium allotrope akin to Graphene and Silicene

M.E. Dávila [1*], L. Xian[2], S. Cahangirov[2], A. Rubio[2*], G. Le Lay[3]

[1]Instituto de Ciencia de Materiales de Madrid-ICMM-CSIC, C/Sor Juana Inés de la Cruz, 3 Cantoblanco 28049-Madrid Spain

[2]Nano-Bio Spectroscopy group, Universidad del País Vasco and European Theoretical Spectroscopy Facility (ETSF), CFM CSIC-UPV/EHU-MPC and DIPC, Av. Tolosa 72. E-20018 San Sebastián, Spain.

[3]Aix-Marseille Université, CNRS, PIIM UMR 7345, 13397 Marseille Cedex, France.

*Corresponding authors: mdavila@icmm.csic.es, angel.rubio@ehu.es.

**Abstract**:

Using a gold (111) surface as a substrate we have grown *in situ* by molecular beam epitaxy an atom-thin, ordered, two-dimensional multi-phase film. Its growth bears strong similarity with the formation of silicene layers on silver (111) templates. One of the phases, forming large domains, as observed in Scanning Tunneling Microscopy, shows a clear, nearly flat, honeycomb structure. Thanks to thorough synchrotron radiation core-level spectroscopy measurements and advanced Density Functional Theory calculations we can identify it to a $\sqrt{3}\times\sqrt{3}R(30°)$ germanene layer in coincidence with a $\sqrt{7}\times\sqrt{7}$ $R(19.1°)$ Au(111) supercell, thence, presenting the first compelling evidence of the birth of a novel synthetic germanium-based cousin of graphene.

**Introduction:**

After the successful synthesis of silicene in 2012 followed by a surge of studies on elemental, novel two-dimensional materials beyond graphene, a daunting quest was to obtain germanene, the germanium based analogue of graphene, already predicted to possibly exist in 2009. Although its fully hydrogenated form, germanane, has been already fabricated by a wet chemistry method in 2013, germanene remained elusive until now. Here we show compelling experimental and theoretical evidence of its first synthesis by dry epitaxial growth on a gold (111) surface.

The discovery of graphene has boosted research in nanoscience on two-dimensional (2D) materials, especially on elemental ones. In 2012, silicene, graphene's silicon cousin [1], has been successfully synthesized on two metallic templates, namely a silver (111) surface [2,3] and the zirconium diboride (0001) surface of a thin film grown on a silicon (111) substrate [4]. One year later, silicene has been also grown on an iridium (111) surface [5]. Germanene, another, germanium based, cousin of graphene, along with silicene, had been predicted to be stable as free standing novel germanium and silicon 2D allotropes in a low buckled honeycomb geometry by Cahangirov et al., already in 2009 [6]. In the quest for germanene, strikingly, its fully hydrogen-terminated partner, germanane (GeH), has been first fabricated from the topochemical deintercalation of the layered van der Waals solid calcium digermanide ($CaGe_2$) [7].

Since silicene, up to now, has been synthesized only in dry conditions under ultra high vacuum (UHV) -with silver (111) as the most favorite substrate-, a tempting way to synthesize germanene would be, seemingly, to try to grow it also on Ag(111) single crystals by germanium molecular beam epitaxy. However, to the best of our knowledge, such a tentative has failed up to now, most probably because germanium prefers to form an ordered $Ag_2Ge$ surface alloy, where Ge atoms, up to a coverage of one-third of a monolayer (1/3 ML), substitute Ag ones at the silver surface. This surface alloy presents a complex "√3x√3" structure [8], which deviates, not only in its geometry but also in its electronic properties [8,9] from the simple √3x√3 reconstruction envisaged earlier [10].

We have thus used, instead, a gold (111) substrate to avoid such a surface alloy formation. Indeed, for silicene synthesis, we deposited silicon on silver (111) surfaces because the inverse system, silver grown on Si(111) surfaces, is well-known to form atomically abrupt interfaces, without intermixing [11]. Our choice of a Au(111) substrate is based on the same strategy. It turns out that among the four noble metals on elemental semiconductor systems studied, namely, Au,Ag/Ge,Si(111) [11], the most similar in several aspects, especially in the growth mode –Stanski-Krastanov (or layer-plus-islands) mode- characterized by the formation of a √3x√3 *R*30° superstructure (or wetting layer) associated with the formation of Au trimers on Ge(111) or Ag ones on Si(111), appeared to be Si/Ag(111) [13] and Ge/Au(111) [14], a trend confirmed in a recent study of Au/Ge(111) [15].

This strategy has paid back. As we will see, we have succeeded in growing a 2D germanium sheet with a *honeycomb* appearance in scanning tunneling microscopy (STM) imaging on top of the Au(111) surface. Its formation bears great similarity with the growth of silicene sheets on the Ag(111) surface [2,3]. We identify it to a germanene sheet after detailed synchrotron radiation spectroscopy measurements of the deposit (Ge 3d) and substrate (Au 4f) shallow core-levels and thanks to advanced Density Functional Theory calculations of the geometry and stability of the system using for the exchange-correlation potentials the General

Gradient Approximation (see supplementary information). Independently, nearly in parallel, another group has concluded to the formation of a germanene layer on a platinum (111) template, appearing, however, strongly distorted, based only on STM observations of a *hexagonal* arrangement and DFT calculations in the basic Local Density Approximation [16].

One can anticipate a major impact of this new discovery because of the expected very high mobilities of the carriers [17], the potential optical applications [18], the predicted robust two-dimensional topological insulator character, nearly up to room temperature, resulting from the large effective spin-orbit coupling [19,20] opening the way to the Quantum Spin Hall Effect [21], the possibility of very high Tc superconductivity [22,23], and, last but not least, the practicability of direct integration in the current electronics industry.

**Results and discussion:**

The methodology we have adopted here, is very similar to the one we used for the synthesis of silicene on Ag(111) (see ref. 2 for details) ; just the silver sample was changed for a gold (111) one and the silicon source in the experimental set-up in Marseille was replaced by a germanium evaporator to deposit Ge atoms onto a clean Au(111) surface prepared in a standard fashion by Ar$^+$ ion bombardment and annealing. The *in situ* cleaned Au(111) surface is characterized by its well-known 22x√3 herringbone structure [24]. Low energy electron diffraction (LEED) and STM observations were performed at room temperature (RT) at different stages of the growth, carried out at several substrates temperatures to determine potential candidates for germanene in an overall multiphase diagram as was already the case for silicon deposition onto Ag(111) [25].

In this article we will focus on one of the phases obtained at ~200°C growth temperature at about 1 ML coverage, as estimated from the 32% attenuation of the Au $4f_{7/2}$ core level intensity. It covers extended regions, larger than 50x50 nm$^2$ in size, with a honeycomb appearance and a very small corrugation of just 0.01 nm, as well as with a weak a long range modulation in STM imaging, as displayed in Fig. 1.

However, as was the case for the main silicene phase on Ag(111) (noted 3x3/4x4 to illustrate the 3x3 reconstruction of silicene in a 4x4 coincidence cell on Ag(111)), here, again, the observed honeycomb arrangement is too large [2] to correspond directly to a germanene primitive cell. Instead, its cell size fits to a √7x√7 *R*(19.1°) superstructure in terms of Au(111), which is in accord with the LEED pattern of Fig. 1b. This somewhat astonishing LEED pattern with many extinctions reflects the diffraction from three main co-existing phases, as schematically shown in Fig. 1c: a √19x√19 *R*(23.4°) phase, a 5x5 one and a √7x√7 *R*(19.1°) phase, the one of prime interest here, noted with reference to the Au(111)1x1 basis vectors. The extra spots noticed at small distance from the integer order ones suggest the presence of a distortion with a long repetition length, most probably reflecting the modulation seen in the STM images, which can be related to underlying remnants of the native herringbone structure of the pristine Au(111) surface. At this stage, we stress that in most circumstances, the growth of silicene on Ag(111) takes place also in several phases, the most frequent LEED pattern corresponding, in this case, to a superposition of diffraction patterns with many extinctions stemming essentially from √13x√13 *R*(13.9°) and 4x4 phases (labeled with reference to Ag(111)1x1), but also, typically, a √19x√19 *R*(23.4°) phase [24]. Hence, somehow, the astonishing LEED pattern we get for the present Ge/Au(111) system should not be such a surprise.

In line with the silicene phases on the Ag(111) surface, the √7x√7 *R*(19.1°) superstructure in terms of Au(111) we focus on here, could be possibly associated to a 2x2 or a √3x√3 *R*(30°) germanene phase on top of the Au(111) surface (symmetry, in each case, imposing a two-domain structure). This view is supported by the synchrotron radiation core-level spectroscopy measurements displayed in fig. 2 (for details on the data acquisition and fitting procedure and parameters see supplementary information).

The clean Au(111) 4f core-levels are fitted with a bulk (B) and a surface shifted component (S) of similar intensities in the highly surface sensitive conditions of the measurements (typically, the kinetic energy of the measured Au $4f_{7/2}$ line at ~46 eV corresponds to the minimum of the escape length of around 0.5 nm). After germanium deposition at ~200°C and the growth of the two-dimensional phases, as displayed in Fig. 1, the total Au 4f intensity at normal emission is reduced by ~32% ; still, a fraction of the Au(111) surface (about 25%), remains uncovered since the surface component is not totally quenched. On the high binding energy side a new component, noted I, representing ~15% of the total Au 4f intensity has developed. The relative intensity ratio I/B at normal emission (0.21) increases to 0.29 at 50° off-normal emission while the S/B ratio barely changes. This testifies that the gold atoms contributing to this I component are at the very top surface. The corresponding Ge 3d core-levels are fitted with a very narrow, asymmetric single component at 50° off-normal emission, in extremely surface sensitive conditions (escape depth estimated at ~0.4 nm) signaling essentially a unique environment of the germanium atoms at the very top surface and their metallic character; we assign the small broad additional component (just 9% of the total intensity) at normal emission to defect sites. The essentially unique Ge species indicate that no formation of a surface alloy occurs, at variance with the $Ag_2Ge$ one formed in the case of Ge deposited on Ag(111) surfaces [8,10] or the one initially formed upon Ge deposition onto the Au(110) surface [26]. In this respect, we note that the formation of a surface alloy is surface depend; typically, it takes place upon Ge deposition onto the Ag(111) surface, but not onto the Ag(110) and Ag(100) ones.

Based on these core-level results and the honeycomb appearance of the STM images of Fig. 1, we can assume that the germanium two-dimensional overlayer grown on top of the Au(111) surface, is composed of germanene sheets arranged either in a √19x√19 *R*(23.4°) supercell (with reference Au(111) 1x1), a 5x5 one and a √7x√7 *R*(19.1°) one. As mentioned above, this last supercell could correspond either to a 2x2 germane reconstructed epitaxial sheet (projected in-plane Ge-Ge distance: dGe-Ge = 0.221 nm) or to a √3x√3 *R*(30°) one (dGe-Ge = 0.255 nm) since the corresponding value for free standing germanene is dGe-Ge = 0.238 nm [6], while the √19x√19 *R*(23.4°) and 5x5 ones could correspond respectively to a 3x3 (projected in-plane Ge-Ge distance: dGe-Ge = 0.242 nm) and a √13x√13 *R*(13.9°) one (dGe-Ge = 0.231 nm).

In the following we address the question of the epitaxial structures for the √7x√7 *R*(19.1°) supercell, since this is the one observed in STM imaging as a honeycomb arrangement. To this end we have undertaken thorough Density Functional Theory calculations to determine the minimum energy configuration within this supercell, even searching for a surface alloy (although very unlikely from the core-level measurements) and also allowing for possible substitution of few Au atoms within the germanene sheet; for details of the calculations, see supplementary information.

The lattice mismatch between the cell sizes of 2x2 free standing germanene and that of the √7x√7 Au (111) surface appears to be small (8.12 Å versus 7.78 Å). However, the 4.2% compression, along with the strong Ge-Au interaction distorts the germanene lattices and induces considerable buckling in the structure. The atomic structures and simulated STM images of the two lowest energy structures for 2x2 germanene on top of the √7x√7 Au supercell are shown in Fig. 3a-b. The average height variations in structures 1 and 2 are 0.150 nm and 0.142 nm, respectively, which are much larger than what is measured in experiments. Therefore, these two structures are not the structures observed in experiments.

On the other hand, a very flat structure is obtained when √3x√3 germanene is placed on top of the √7x√7 Au surface. The lowest energy structure in this configuration is shown in Fig. 3c. The height variations of different Ge atoms are less than 0.05 nm. Compared with the previous two structures; this one has a lower absorption energy, which is defined as:

$$E^{abs} = (E_{Ge/Au(111)} - E_{Au(111)})/N_{Ge} - E_{Ge}, \qquad (1)$$

where $E_{Ge/Au(111)}$, $E_{Au(111)}$, and $E_{Ge}$ represent the total energies of the germanene-covered Au slab, the pure Au slab, and the isolated Ge atom, respectively, and $N_{Ge}$ represents the number of Ge atoms in the supercell. As seen in Table I, the absorption energy of this structure is even lower than the bulk cohesive energy of diamond Ge, indicating that it is more energetically favorable to form such a layer structure than to form Ge clusters on the Au(111) surface. The simulated STM image for structure 3 is shown in Fig. 3c. As highlighted with blue circles, there is a darker region for each supercell. With the consideration of the tip effects, an image similar to those observed in experiments may be obtained by Gaussian smearing with width σ = 0.6 Å, as shown in Fig. 3d.

The Au 4f surface core level shift calculated for a pure 7 layer Au slab is -0.35 eV, which agree well with the experimental value. The Ge 3d core level shift between different Ge atoms in structure 3 has three components, as shown in table I, with the energy of the main component set to 0.00 eV. The intensity ratio between these three components is 1:4:1. The difference between each small component and the main component is less than 0.09 eV, which is also consistent with our experimental results. The calculated Au 4f core level shift for the Au atoms below Ge atoms in structure 3 has two components: one is shifted by 0.37 eV, the other is shifted by 0.15 eV, with relative intensity ratio 4:3. These two components, the signal of which are attenuated by the germanene layer above, can be related to the component I, the interface component, located at the left of the bulk peak in experiments.

Although our calculations on structure 3 agrees very well with our experiments we would like to explore also the possibility of forming a Ge-Au surface alloy on the Au(111) surface. For example, switching the position of one Ge atom and one Au atom in structure 3, the total energy of the system can be even lower. The lowest energy structure by switching one pair of atoms is shown in Fig. 4a. The total energy of this structure (named as structure 4) in the √7x√7 Au supercell is lower than that of the structure 3 by 0.43 eV. A larger scale calculation in a 2√7x2√7 Au supercell shows that the total energy decreases almost linearly with the increasing number of switching pairs, from 1 pair up to 3 pairs. The surface of such a structure is also very flat with height variations less than 0.3 Å. The simulated STM image is shown in the bottom panel of Fig. 4a. Calculations indicate that the Ge 3d core level spectra should have three components (see Table I). In particular, the two surface components are separated by about 0.17 eV, and the component corresponding to the subsurface Ge atom shifts to 0.34 eV higher than the majority of surface components. The disagreement between the calculation and the experimental results for the Ge 3d core-level indicates that structure 4 is not the structure observed in experiments.

When there are only five Ge atoms on the Au(111) surface within the √7 x √7 Au supercell, the relaxation of all the surface layers will result in a structure similar to that shown in Fig. 4b , i.e., one Au atom is pulled out of the Au surface and forms a honeycomb lattice with the other Ge atoms, leaving a vacancy in the subsurface (the locations of the vacancy are highlighted with red circles). Following the definition in Eq. (1), the absorption energy per Ge of this structure is even lower than that of structure 3 by 0.049 eV. However, with a vacancy in the subsurface, this structure is not a stable structure. Other Au atoms will diffuse from the bare surface or from the bulk to fill in the vacancy to form a more stable structure. The new structure consists of a honeycomb germanene lattice with some Ge atoms substituted by Au atoms and the original Au surface. The lowest energy structures with one and two Ge atoms substituted by Au atoms are as shown in Fig. 4b and c, with the vacancies in Fig. 4b filled with Au atoms. However, it is not possible to compare the absorption energy of Ge atoms in structure 3 with that of structure 5, since Eq. (1) does not apply to the latter. Instead we introduce a new definition for absorption energy of Ge atoms in systems having Au atom substitutions like structure 5:

$$E^{abs} = (E_{Ge+Au/Au(111)} - E_{Au(111)} - N_{Au}E_{bulk\ Au})/N_{Ge} - E_{Ge}, \qquad (2)$$

where $N_{Au}$ is the number of substituting Au atoms and $E_{bulk\ Au}$ is cohesive energy per atom of bulk gold. After introducing Au atoms in the honeycomb lattice, the Ge-Au layer is still very flat with height variations less than 0.5 Å. The simulated STM images are shown in the lower panels of Fig. 4b and c. For structure 5, there is a darker region in every supercell. For structure 6, besides a darker region, there is a brighter spot due to the Au atom in each supercell. With proper smearing, images similar to those observed in experiments may be acquired.

To conclude on these theoretical results, our DFT calculations show that the atomic model composed of a √3x√3 reconstructed germanene sheet on top of a √7x√7 Au(111) surface matches the STM observations and the core-level measurements very well, although the germanene layer may possess some gold atom substitutions.

**Conclusions:**

To summarize, a two-dimensional germanium layer, forming several phases, has been grown *in situ* by dry deposition on the Au(111) surface, similarly to the formation of silicene on Ag(111). One of these phases displays a clear honeycomb structure with a very weak corrugation in STM imaging. Detailed core-level spectroscopy measurements along with advanced DFT calculations allow us to identify this phase to a √3x√3 reconstructed germanene layer on top of a √7x√7 Au(111) surface. By this, we provide compelling evidence of the birth of nearly flat germanene, a novel synthetic germanium allotrope, which does not exist in nature, which is a new cousin of graphene.

**Acknowledgments:**

Drs F. Cheynis, D. Chiappe, A. Ranguis and A. Resta, are warmly acknowledged for their contributions in the measurements. We are grateful to Dr. M. Radović and Dr. R. S. Dhaka at Swiss Light Source and I. Vobornik at the Italian synchrotron radiation source Elettra, whose outstanding efforts have made these experiments possible. Funding from the 2D-NANOLATTICES project within the 7th Framework Programme for Research of the European Commission, under FET-Open grant number 270749 has been greatly appreciated. We acknowledge support by the European Research Council Advanced Grant DYNamo (ERC-2010-AdG-267374), Grupos Consolidados UPV/EHU del Gobierno Vasco (IT-578-13) and European Commission project CRONOS (Grant number 280879- 2).

# Supplementary information

### I.   CORE-LEVEL SPECTROSCOPY

Here, we present experimental details during the high-resolution Scanning Tunneling Microscopy and synchrotron radiation core-level spectroscopy studies of Germanium grown on Au(111) in situ.

Samples were prepared *in situ* in two separate ultrahigh vacuum (UHV) systems, i.e. one with Low Energy Electron Diffraction and STM and one with LEED and Angle Resolved PhotoElectron Spectroscopy (ARPES). STM images were recorded at room temperature using an Omicron variable temperature STM at Aix-Marseille University and the ARPES experiments were first performed on the Surface/Interface Spectroscopy (SIS) X09LA beamline at the Swiss Light Source, Paul Scherrer Institut, Villigen, Switzerland, then repeated (with confirmation) at the APE beamline of the Italian synchrotron radiation facility, Elettra in Trieste. At the SLS (data presented here), the beamline was set to Linear polarized light with a photon energy, hv = 135 eV with an energy resolution of 80 meV, and data were acquired at Room Temperature using a VG-Scienta R4000 electron analyzer. The binding-energy scale was calibrated with a copper reference sample in direct electrical and thermal contact with the film. The base pressure of the UHV systems was below $5 \times 10^{-11}$ mbar during the entire measurement and no sign of sample and/or data quality degradation was observed. Our results were reproduced on several occasions, using different samples grown under the same conditions. A single crystal Au(111) substrate was cleaned in vacuum by 1.5-keV Ar ion sputtering for 30 min at $5 \times 10^{-5}$ mbar. Subsequently, annealing in vacuum at 500° C for 30 min was performed to cure Ar ion sputtering damage and obtain flat and well-ordered surface. The annealing-sputtering cycle was repeated as many times as necessary to obtain a clean surface free of C and O contaminants as verified by *in situ* x-ray photoelectron spectroscopy (XPS). Here, slightly less than 1 ML of Ge was deposited on the substrate at 200° C by a resistance heated crucible resulting in sharp LEED patterns. A series of experiments was undertaken to determine the rate of germanium evaporation in vacuum as a function of temperature.

The Au 4f and Ge 3d core-levels (CLs) have been fitted using standard methods with the following parameters:

Clean Au(111) Normal Emission (NE) : Au 4f
Areas: B 15,02 ; S 15.83
Spin-orbit split: 3.678 eV,
Branching ratio: 0.6,
Gaussian widths: 121 (S) and 200 (B) meV,
Lorentzian width: 300 meV.
The asymmetry parameter of the Doniach-Sunjic line Profile is $\alpha = 0.012$,
The energy difference between the two components Surface and bulk is 0.31 eV.

Ge/Au(111) NE : Au 4f
Areas: B 14,9 ; S 3,12 ; I 3,11
Branching ratio: 0.6,
Gaussian widths: (B) 204 ; (S) 294 ; (I) 311 meV,
Lorentzian width: 300 meV.
$\alpha = 0.012$,
The energy difference between the two components Surface and Bulk is 0.31 eV and B to I: 0.3 meV

Ge/Au(111) 50° Off Normal: Au 4f
Areas: B 6.54 ; S 1.4 ; I 1.9
Spin-orbit split: 3.678 eV,
Branching ratio: 0.58,
Gaussian widths: 180 (S) ; 150 (B) ; 200 (I) meV,
Lorentzian width: 300 meV.
$\alpha = 0.012$ The energy difference between the two components Surface and bulk is 0.31 eV and (B) to (I) 0.3 eV

Ge/Au(111) NE : Ge 3d
Areas, Peak1: 1.80713 ; Peak2: 0.18152
Spin-orbit split: 0.551 eV
Branching Ratio: 0.69
Gaussian widths: 112 and 370 meV, respectively,

Lorentzian width: 150 meV.

α = 0.12,

Energy difference between the two components is 0.245 eV.

Ge/Au(111) 50° Off Normal: Ge 3d

Areas: Peak1: 1.7816; Peak2 : 0.00

Branching ratio: 0.65,

Gaussian widths: 114 and 370 meV, respectively.

## II. COMPUTATIONAL METHODS

We have performed first principle calculations based on density functional theory (DFT) as implemented in the VASP code [1]. The exchange-correlation potentials are treated with the generalized gradient approximation (GGA) of Perdew, Burke and Ernzerhof (PBE) [2]. Ionic cores of atoms are modeled with the projector-augmented wave (PAW) pseudopotential method [3]. Plane wave basis set with a energy cutoff of 310 eV are employed for the valence electron wave functions. The calculations are performed in a √7x√7 Au supercell on the Au(111) surface with theoretical optimized bulk Au-Au distance 0.294 nm. The supercell Brillouin zone is sampled with 9 x 9 x1 k-point grids. The Au surface is modeled with a 6-layer slab and a vacuum region larger than 1.6 nm in the z direction perpendicular to the surface. The bottom three Au layers are kept fixed, while all the other atoms are relaxed in the calculations until all forces are converged within 0.01 eV/Å$^{-1}$. Dipole correction is added along the z direction to eliminate the artificial long range interaction between periodic slab images [4]. STM simulations are performed base on the Tersoff-Hamann model [5]. Core level shift for the Au 4f and Ge 3d electrons are calculated within the final state approximation [6]. Considering the lattice constant for germanene and Au surface, we have examined several possible candidates for monolayer germanene on top of the Au (111) surface, including 2 x 2 and √3 x√3 germanene on top of the √7 x √7 Au surface.

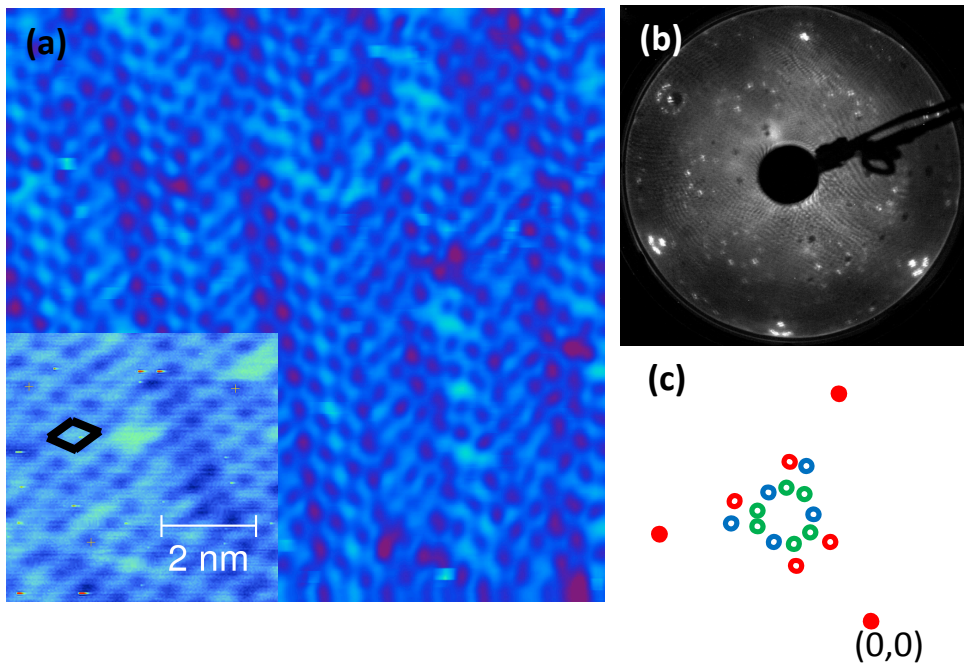

**Fig. 1** (a) 16.2 nm x 16.2 nm STM image of the modulated honeycomb √7x√7 superstructure with a zoom-in at the bottom left corner (-1.12 V, 1.58 nA; the √7x√7 unit cell is drawn in black); (b) associated LEED pattern taken at 59 V; schematic illustration of one sixth of the pattern, filled dots: hidden (0,0) spot and integer order spots, open circles: spots corresponding to the √7x√7 superstructure (in red), the √19x√19 one (in green) and the 5x5 (in blue).

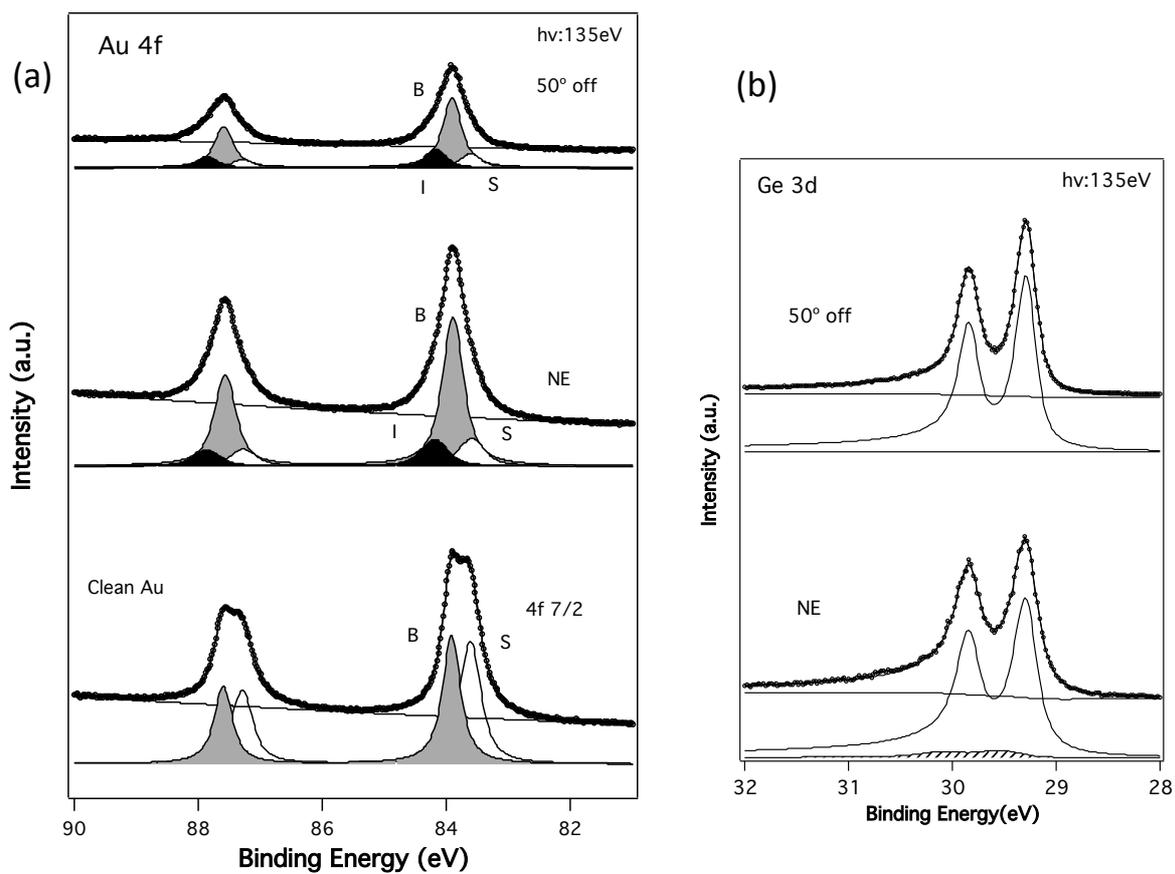

**Fig. 2** Synchrotron radiation Au 4f (left) and Ge 3d (right) core-level spectroscopy measurements at normal (NE) and 50° off normal emission, taken at hν = 135 eV for the two-dimensional phases of Ge grown on Au(111) at ~200°C; B, S and I are bulk, surface and interface components, respectively.

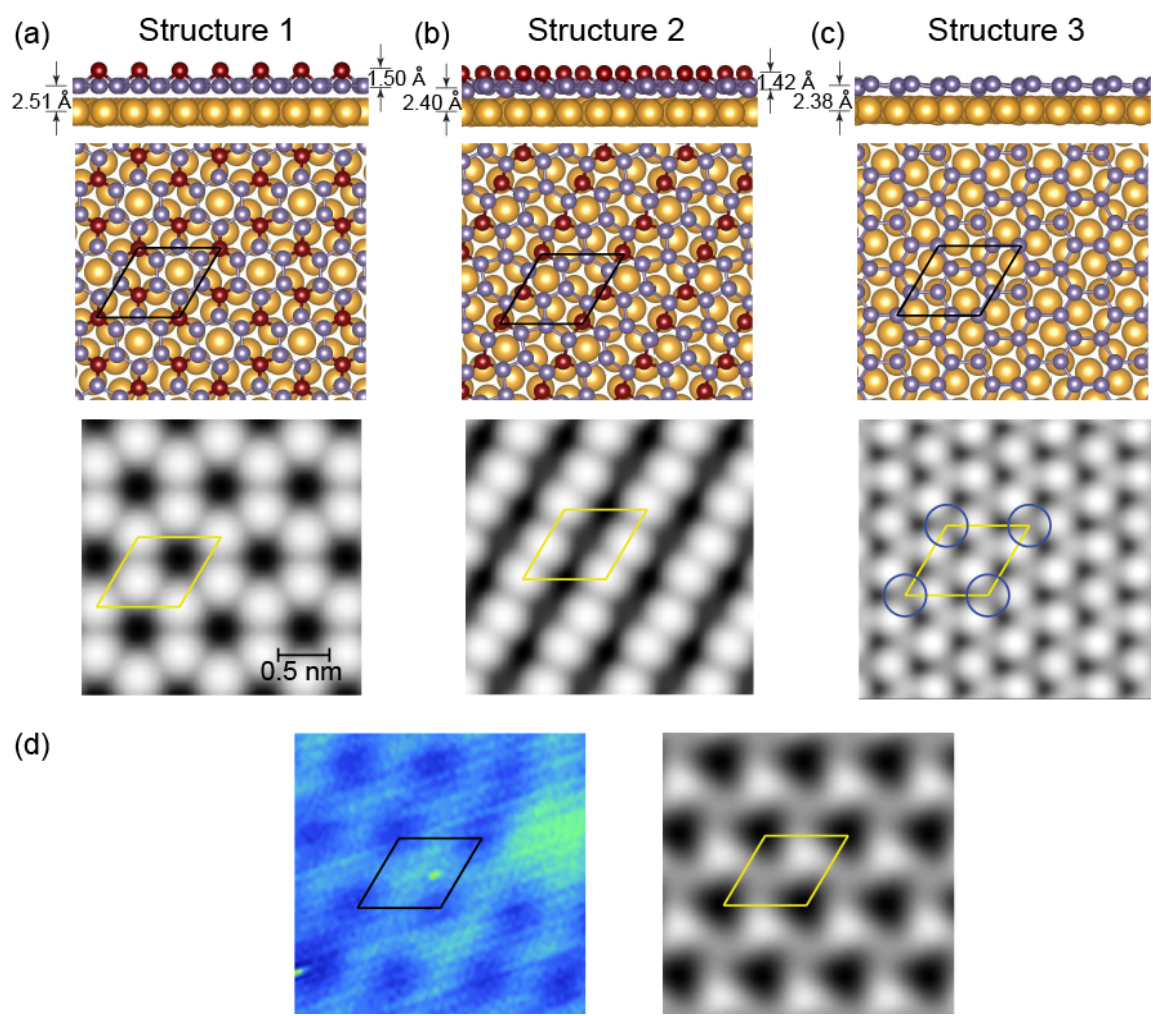

**Fig. 3** (a)-(c) Atomic structures and simulated STM images of three different models of germanene on the √7x√7 Au(111) surface. Structures 1 and 2 have 2x2 periodicity while structure 3 has √3x√3 periodicity with respect to germanene. The protruding Ge atoms are highlighted in dark red. The supercells in the STM images are highlighted with yellow lines. (d) Comparison between the experimental image (left panel) and the simulated STM image for structure 3 after smearing (right panel).

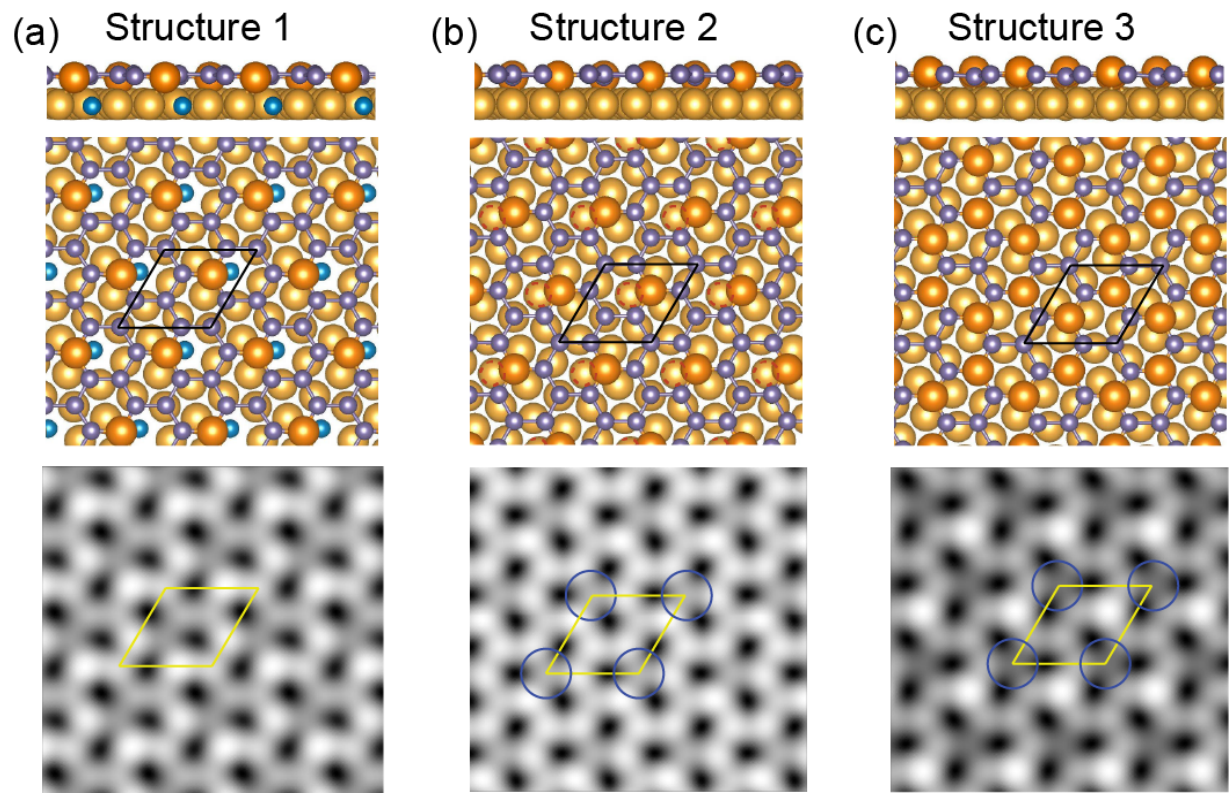

**Fig. 4** Atomic structure and simulated STM images of three different germanene structures having substitutional Au atoms. The Au atoms in the surface layer are highlighted in orange and the Ge atoms in the subsurface layer are highlighted in light blue.

|  | Energy per Ge atom (eV/atom) | Buckling (Å) | Ge 3d core level shifts (eV) | Au 4f core level shifts (eV) |
| --- | --- | --- | --- | --- |
| Structure 1 | -3.641 | 1.50 |  |  |
| Structure 2 | -3.648 | 1.42 |  |  |
| Structure 3 | -3.744 | 0.47 | -0.09, 0.00, 0.06 | 0.15, 0.37 |
| Structure 4 | -3.815 | 0.28 | 0.00, 0.17, 0.34 | 0.24, 0.36, 0.43 |
| Structure 5 | -3.847 | 0.23 | -0.14, 0.00, 0.11 | -0.11, 0.27, 0.33 |
| Structure 6 | -3.868 | 0.41 | ≤0.03 | -0.12, 0.17, 0.22, 0.29 |

**TABLE I**: Absorption energies and core level shifts for different germanene structures on Au (111) surface. The energy of bulk germanium in its cubic diamond-type structure is calculated to be 3.727 eV/atom. The main components of the Ge 3d core levels are set to 0.00eV.